\documentstyle[12pt]{article}
\def\mathrm#1{{\rm #1}}

\newcommand{\be}{\begin{equation}}
\newcommand{\ee}{\end{equation}}

\def\psnormal{\textwidth=16cm\textheight=21.5cm
          \oddsidemargin=0.5cm\evensidemargin=0cm
          \topmargin=0cm\parindent=1cm}
\psnormal

\begin{document}
\pagestyle{empty}

\hspace{3cm}

\vspace{-3.4cm}
\rightline{{ FTUAM 93/22}}

\vspace{1.0cm}
\begin{center}
{\bf NON--UNIVERSAL SOFT SCALAR MASSES IN SUPERSYMMETRIC THEORIES}
\vspace{0.8cm}

A. LLEYDA and C. MU\~NOZ${}^{*}$
\vspace{0.8cm}

 Dept. de F\'{\i}sica Te\'{o}rica C-XI, \\
Univ. Aut\'{o}noma de Madrid, E-28049 Madrid, Spain
\vspace{1.0cm}

\end{center}

\centerline{\bf Abstract}
\vspace{0.3cm}

\noindent
The existence of non--universal soft masses is the most general situation in
supersymmetric theories. We study the consecuences that this
situation has for the low--energy sparticle spectrum. In particular,
we analize in detail the contribution to the scalar mass
renormalization group equations of the $U(1)_Y$ D--term. We obtain
analytic expressions for the evolution of masses of the three
generations and these allow us to show that such a contribution can
produce important modifications on the spectrum. The necessity to avoid
flavour changing neutral currents does not constrain this result.
Finally, we discuss a realistic example in the context of string
theory where the departure from universality is large.

\vspace{1.5cm}
\begin{flushleft}
{FTUAM 93/22} \\
{July 1993}
\end{flushleft}
\vspace{3cm}
${}^{*}$ bitnet address: kmunoz@vm1.sdi.uam.es

\psnormal

\newpage
\pagestyle{plain}
\pagenumbering{arabic}

Supersymmetry (SUSY) is very important in order to solve the gauge
hierarchy problem and to unify all fundamental interactions. The
simplest supersymmetric extension of the Standard Model is the
so--called Minimal Supersymmetric Standard Model (MSSM). It is an
effective low--energy supergravity (SUGRA) theory with the Standard
Model gauge group, global $N=1$ SUSY broken softly and the minimal
particle content (i.e. three generations of quark and lepton superfields
plus two Higgs doublets, $H_1$ and $H_2$, of opposite hypercharge as
well as the gauge superfields) \cite {hpn}. In particular, the soft
scalar masses generated after SUSY--breaking in the effective observable
scalar potential are of the type:
\begin{eqnarray}
V_{eff}= \sum_{\phi} m_{\phi}^2 \; | \phi |^2
\label{veff}
\end{eqnarray}
The rich spectrum of sparticles of the MSSM has been studied almost
exclusively under the assumption\footnote{Common gaugino masses are also
usually considered in the analyses of the MSSM. See ref.\cite{bfmo} for
exceptions, where the effect of relaxing this assumption on neutralinos
is studied.} $m_{\phi}=m$, i.e. the scalars have universal masses at the
unification scale ($ M_{GUT} $). In this way, after renormalizing the
theory, all
the physical masses can be calculated in terms of a small number of soft
parameters and therefore the analysis is simplified. If, in the SUGRA
theory under consideration, the kinetic terms of the fields are canonical
then the above assumption is true and $m$ is the gravitino mass. However,
for a general K\"{a}hler manifold, the relation $m_{\phi}=m$ is not
fulfilled and non--universal scalar masses can emerge:
\begin{eqnarray}
m_{\phi} = c_{\phi} \; m
\label{mpcpm}
\end{eqnarray}
where $c_{\phi}$ are constants whose values will depend on the specific
mechanism of SUSY--breaking. Now the analysis of the sparticle spectrum
is more difficult since new soft parameters must be taken into account.
In fact, this is what happens in the context of SUGRA theories coming
from superstrings, where the soft masses of the scalar fields can be
explicitly calculated and they show a lack of universality [3-5] unlike the
usual assumption of the MSSM.

In this paper we study the
consecuences that the departure from the common assumption of
universal soft scalar masses has for the low--energy SUSY spectrum. In
particular, we analize in detail the contribution to the scalar mass
renormalization group equations (RGEs) of the $U(1)_Y$ D--term (see
eq.(\ref{dmdt})). We
give analytic expressions for the evolution of masses of the three
generations and these allow us to show that such a contribution can
produce important modifications on the spectrum. The necessity to
avoid flavour changing neutral currents (FCNC) does not constrain this
result. Finally, we discuss a realistic example in the context of
string theory where the departure from universality is large.

Let us consider for the moment the case of first-- and
second--generation sparticles. To calculate the soft SUSY breaking
masses at low energies from the parameters at the unification scale, we
have to use the RGEs [6-9]. Although
Yukawa couplings contribute in the RGEs, their small effects can be
safely neglected unlike the case of the third generation, where top,
bottom, and tau Yukawa couplings can be large. Then
\begin{eqnarray}
{\frac{d m_{\phi}^2}{d t}}=\sum_{i=1}^3 {\frac{\alpha_i}{\pi}}
C_i^{\phi} M_i^2 - {\frac{\alpha_1}{4 \pi}} Y_{\phi} S \; ,
\label{dmdt}
\end{eqnarray}
where $t\equiv \ln{\frac{M_{GUT}^2}{Q^2}}$ , $M_i$ are the gaugino
masses, $Y_{\phi} $ are the hypercharges, $C_i^{\phi} $ is the quadratic
Casimir corresponding to each scalar ( $C=\frac{N^2-1}{2N}$ for $SU(N)$,
$C = Y^2$ for $U(1)_Y$) and the gauge coupling constants at $M_{GUT}$
verify
${\alpha}_3(0)={\alpha}_2(0)=\frac{5}{3}{\alpha}_1(0)={\alpha}_{GUT}$.
Notice that in these equations there is a contribution to the scalar
masses coming from the $U(1)_Y$ D--term\footnote{This term was also
included in a different context in the analysis of ref.\cite{hk}. The
authors studied the phenomenological
consecuences of spontaneous breaking of intermediate
scale gauge symmetries induced by universal soft terms.} parametrized
in the general case by:
\begin{eqnarray}
S =  \sum_{\phi} d(R_{\phi}) Y_{\phi} m_{\phi}^2 \; ,
\label{sdry}
\end{eqnarray}
where $d(R_{\phi})$ is the dimension of the representation $R$
associated to the $\phi$ scalar. The evolution of $S$ is given by
\cite{apw}:
\begin{eqnarray}
\frac{dS}{dt} =- \frac{\alpha_1}{4 \pi} \big( \ \sum_{\phi}
d(R_{\phi})Y_{\phi}^2\ \big) \ S \; .
\label{dsdt}
\end{eqnarray}
For the MSSM eqs.(\ref{sdry},\ref{dsdt}) read
\begin{eqnarray}
S = m_2^2 - m_1^2 \;\; +  \displaystyle \sum_{generations}
( m_{Q_L}^2+m_{D_R}^2+m_{E_R}^2-m_{L_L}^2-2 m_{U_R}^2 ) \; ,
\label{smmm}
\end{eqnarray}
\begin{eqnarray}
\frac{dS}{dt}= - \frac{\alpha_1}{4 \pi} b_1 S \; ,
\label{dsdtalfa}
\end{eqnarray}
where $U_R$, $D_R$ ($Q_L=(U_L,D_L)$) stand for the right--handed
(left--handed) squarks, $E_R$ ($L_L=(V_L,E_L)$) for the right--handed
(left--handed) sleptons and $b_1=11$ is the one--loop coefficient of the
beta function calculated using the following hypercharges:
$6Y_{Q_L}=3Y_{D_R}=Y_{E_R}=-2Y_{L_L}=-\frac{3}{2}Y_{U_R}=1$.
Finally, $m_{1,2}^2 \equiv m_{H_{1,2}}^2+{\mu}^2$, where ${\mu}$ is the
supersymmetric mass term in the superpotential ${\mu} H_1 H_2$. The
solution of eq.(\ref{dsdtalfa})
\begin{eqnarray}
S(t)= \frac{S(0)}{1+ \frac{\alpha_1 (0)}{4 \pi} b_1 t} \;\; ,
\label{stso}
\end{eqnarray}
implies that if $S$ vanishes at some scale, it vanishes everywhere. For
the case of universal boundary conditions for the scalar masses
$m_{\phi}(0)=m$ , $S(0)=0$ (see eq.(\ref{smmm})). However in the most
general case of non--universality, $S(0) \neq 0$ and therefore $S$ must
be included in the RGEs (\ref{dmdt}). We will see below that although
this term in eq.(\ref{dmdt}) is multiplied by $\alpha_1$, its influence
can still be sufficiently large as to modify substantially the
low--energy scalar masses. It is worth noticing here that although the
necessity to avoid large FCNC constrains the amount of non--universality
and suggests that, in a viable model, d-- and s--squarks are nearly
degenerate at $M_{GUT}$, and likewise for the u-- and c--squarks and the
selectron and smuon, different scalars within the same generation can
have different masses. This allows the second term in eq.(\ref{smmm}) to
be large. Besides, the first term depends on the difference between the
masses of the Higgses and therefore is not constrain by FCNC either.

Eq.(\ref{dmdt}) can be analytically solved and for the MSSM we obtain:
\begin{eqnarray}
m_{\phi}^2(t) = m_{\phi}^2(0) + \sum_{i=1}^3 M_i^2(0) f_i^{\phi}(t) -
Y_{\phi} S(0) {f'_1}(t) + M_Z^2 \big( {T_3}_L^{\phi} {cos}^2 \theta_W -
Y_{\phi} {sin}^2 {\theta}_W \big) cos2 \beta
\label{mcmm}
\end{eqnarray}
where
\begin{eqnarray}
&&f_i^{\phi}(t)=\frac{2C_i^{\phi}}{b_i}\Big[ 1- \frac{1}{(1+
\frac{\alpha_i(0)}{4\pi}b_it)^2}\Big] \; , \; f'_1(t)=\frac{1}{b_1}
\Big[ 1- \frac{1}{(1+\frac{\alpha_1(0)}{4\pi}b_1 t)} \Big] \; ,
\nonumber\\
&&tan \beta = \frac{\langle{H_2}\rangle}{\langle{H_1}\rangle} \;\; ;
\;\; b_3 = -3 \; , \; b_2 = 1 \; , \; b_1 =11 \; .
\label{ftcb}
\end{eqnarray}
In eq.(\ref{mcmm}) we have already included the contribution of the
scalar potential D--terms to the scalar masses after the Higgs bosons
get vacuum expectation values (VEVs). One has to add to eq.(\ref{mcmm})
the $(mass)^2$ of the corresponding fermionic partner. The $f's$
evaluated at $M_Z$ are\footnote{Notice to obtain the physical scalar
masses in eq.(\ref{mcmm}) the RGEs must be integrated to
$t=\ln{\frac{M_{GUT}^2}{m_{\phi}^2}}$. We will take into account this
subtleness in the figures below.}:
\begin{eqnarray}
&&f'_1(M_Z)=0.054 \;\; , \; f_3^{\phi}(M_Z)=7.093 \;\; , \;
f_2^{\phi}(M_Z)=0.496 \;\; , \; f_1^{\phi}(M_Z)=0.152\;Y_{\phi}^2 \;\; ,
\label{f1mz}
\end{eqnarray}
where the different ${\alpha}_i(0)$ were computed by running up to
$M_{GUT}$ the three gauge coupling constants obtained from the following
experimental inputs:
\begin{eqnarray}
M_Z=91.175 \; GeV \; , \; {\alpha}_3(M_Z)=0.125 \; , \;
{\alpha}_{em}(M_Z)=\frac{1}{127.9} \;\; , \; {sin}^2{\theta}_W(M_Z)=0.23
\label{mz91}
\end{eqnarray}
As discussed above, in general, $m_{\phi}(0)=c_{\phi} m$ so it is more
useful for us to write eq.(\ref{mcmm}) as
\begin{eqnarray}
m_{\phi}^2(t) = \big( c_{\phi}^2-c^2Y_{\phi}\;f'_1(t) \big) \; m^2 +
\sum_{i=1}^3 M_i^2(0) f_i^{\phi}(t) +
M_Z^2 \big( {T_3}_L^{\phi}{cos}^2\theta_W -
Y_{\phi}{sin}^2{\theta}_W \big) cos2 \beta
\label{mtcc}
\end{eqnarray}
where the influence of the $U(1)_Y$ D--term becomes apparent as a
modification of the soft masses at $M_{GUT}$. The value of $c^2$ is
defined as (see eq.(\ref{smmm})):
\begin{eqnarray}
c^2 \equiv c_2^2 - c_1^2 \; + \sum_{generations}
(c_{Q_L}^2 + c_{D_R}^2 + c_{E_R}^2 - c_{L_L}^2 - 2 c_{U_R}^2)\; .
\label{cdcd}
\end{eqnarray}
Notice that $c_{\phi}^2$ must be a positive number, otherwise the mass
squared of the scalars would already become negative at $M_{GUT}$.
However, let us remark that $c^2$ can be a negative number. Finally,
the universal case is given by $c^2=0$.

Let us now discuss the importance that the $U(1)_Y$ D--term contribution
to the RGEs has for the low--energy spectrum. Notice to neglect this
contribution the condition $f'_1(t)\;|\;c^2\;Y_{\phi}\;| \ll c_{\phi}^2$
must be fulfilled. But this is by no means the most general situation
since $|c^2|$ can be sufficiently large as to compensate the small value
of $f'_1$ (see eq.(\ref{f1mz})). It is very easy to build examples of
this type. For instance $m_{H_1}(0)=m_{H_2}(0)=m$ and $m_{L_L}(0) =
m_{E_R}(0)=m_{U_R}(0)=\frac{1}{2}m_{D_R}(0)=\frac{1}{2}m_{Q_L}(0) = m$
for the three generations of particles give rise to $c^2=18$ and
therefore $f'_1(t) \; | \; c^2 \; Y_{\phi} \; | \; \sim c_{\phi}^2$  for
$\phi = L_L,E_R,U_R$. In order to see the numerical importance that the
$U(1)_Y$ D--term contribution can have in the computation of low--energy
scalar masses, let us consider the previous example for the $E_R$--type
sleptons and $U_R$--type squarks taking as input soft parameters $m=150$
GeV and $M_i(0)=50$ GeV  $i=1,2,3$. Then, for $c^2=18$ and taking for
simplicity a vanishing scalar potential D--term (i.e. $tan \beta = 1$)
we obtain the following masses: $m_{E_R}(M_Z)=32$ GeV and $m_{U_R}(M_Z)=
234$ GeV. However, the universal case $c^2=0$ implies $m_{E_R}(M_Z)=151$
GeV and $m_{U_R}(M_Z)=201$ GeV. This particular example shows that the
value of $c^2$, which determines the departure from universality of soft
scalar masses, can be crucial to determine the correct supersymmetric
spectrum. The larger is $|c^2|$ the more important will be the $U(1)_Y$
D--term.

A more complete analysis valid for any value of the soft
parameters m, M is shown in figs.1 and 2 for the right selectron and the
right U-squark respectively. We take for simplicity $ m_{E_R}(0) =
m_{U_R}(0) = m$ and $M_i(0)=M \;\; i=1,2,3$. It is straightforward to
extend this study for the other particles and non--universal gaugino
masses using eq.(\ref{mtcc}). The solid (dotted) lines in the figures
show the contours in the m, M
plane corresponding to different physical masses\footnote{We omit to
plot the contours for negative values of M since the figures are
symmetrical as we will explicitly show below.} in the limit
$tan \beta = 1$ ($tan \beta = \infty $). For squark masses both types of
lines are almost indistinguishable since the contribution of the last
term of eq.(\ref{mtcc}) is practically negligible compared to the other
terms. Slepton masses are more sensitive to this contribution as can be
seen in fig.1. Different posible values of $c^2$ are considered in order
to study the modifications that they introduce in the spectrum in
comparison with the universal case $c^2=0$. The curves in the figures
are conics since eq.(\ref{mtcc}) can be written as
\begin{eqnarray}
\frac{M^2}{a^2} + \frac{m^2}{b^2} = 1
\end{eqnarray}
with
\begin{eqnarray}
a^2 &\equiv& \frac{m_{\phi}^2(t) -
M_Z^2 \big({T_3}_L^{\phi}{cos}^2\theta_W-Y_{\phi}{sin}^2{\theta}_W\big)
cos2 \beta} { \displaystyle \sum_{i=1}^3 f_i^{\phi}(t)}
\nonumber\\
b^2 &\equiv& \frac{m_{\phi}^2(t) -
M_Z^2 \big({T_3}_L^{\phi}{cos}^2\theta_W-Y_{\phi}{sin}^2{\theta}_W\big)
cos2 \beta} {c_{\phi}^2-c^2 Y_{\phi} f'_1 (t)}
\end{eqnarray}
For fixed values of the low--energy scalar masses $a^2$ is a constant
whereas $b^2$ is a function of $c^2$ (as can be seen in the figures).
When the condition $ c_{\phi}^2 - c^2 Y_{\phi} f'_1(t) > 0 $ ($<0$) is
fulfilled the curves are ellipses (hyperbolas).

As mentioned above, it is obvious from the figures that the value of
$c^2$ is very important in order to determine the correct supersymmetric
spectrum at low--energies. A negative value of $c^2$ will increase the
mass of the right selectron with respect to the universal case $c^2=0$.
On the contrary, a positive value of $c^2$ will decrease its mass and
for values $c^2 \stackrel{>}{{}_{\sim}} 20$ the ellipses are transformed
in hyperbolas. In the latter case the region of soft parameters m, M
excluded by the present experimental bounds ( $m_l > 45$ GeV ) may be
enormously increased with respect to the universal case $c^2=0$, as can
be seen in fig.1. For instance a bound $|M|>115$ GeV is obtained for
$tan \beta =1 $. For the right U--squark the analysis will be the
opposite since its hypercharge $Y_{U_R}=-\frac{2}{3}$,
which appears in the first
term of eq.(\ref{mtcc}), has the opposite sign to the one of the right
selectron $Y_{E_R}=1$. In fig.2 we see that a negative (positive) value of
$c^2$ will decrease (increase) its mass with respect to the universal
case $c^2=0$. Notice that the influence of the $U(1)_Y$ D--term,
although significant, is smaller than in the right selectron case (e.g.
the ellipses are transformed in hyperbolas for $|c^2| \simeq 30$) due to
the fact that $|Y_{U_R}|<|Y_{E_R}|$ and the contribution of the second
term of eq.(\ref{mtcc}) is more important for the right U--squark.
Finally, let us remark that a similar analysis can be carried out for
the other particles of the first and second generation. In
ref.\cite{fhkn} a program where measurements of three sparticle masses at
future accelerators determine, in the universal case, the values of $tan
\beta$, m, M was proposed. It was also remark that, if additional
sparticle masses are not consistent with the obtained values
of those parameters, that could be
a signal of new physics. In our approach that would be a signal of
non--universality of soft terms and the above analysis would have to be
applied.

Let us now consider the case of third--generation sparticles. The RGEs
are more complicated than in the previous case since Yukawa couplings
are present. However, as it is well known, in the interesting
aproximation where all Yukawa couplings except the one of the top $h_t$
are neglected, the formulae are simplified quite a lot and one can give
some analytic expressions for the evolution of masses. In this
aproximation the above analysis still applies for all the sleptons and
squarks of the third generation except stops and left sbottom. The RGEs
for the latter are:
\begin{eqnarray}
&&\frac{dm_{Q_L}^2}{dt}=\big(\frac{16}{3} \frac{\alpha_3}{4 \pi} M_3^2+
3\frac{\alpha_2}{4\pi}M_2^2+\frac{1}{9}\frac{\alpha_1}{4\pi}M_1^2\big)-
\frac{h_t^2}{(4 \pi)^2} (m_{Q_L}^2+m_{U_R}^2+m_2^2+A_t^2-{\mu}^2) -
\frac{1}{6} \frac{\alpha_1}{4\pi} S
\nonumber\\
&&\frac{dm_{U_R}^2}{dt}=\big(\frac{16}{3} \frac{\alpha_3}{4 \pi} M_3^2+
\frac{16}{9} \frac{\alpha_1}{4 \pi} M_1^2 \big)-
2 \frac{h_t^2}{(4 \pi)^2} (m_{Q_L}^2+m_{U_R}^2+m_2^2+A_t^2-{\mu}^2 )+
\frac{2}{3} \frac{\alpha_1}{4 \pi} S
\nonumber\\
&&\frac{dm_1^2}{dt}=(3\frac{\alpha_2}{4\pi}M_2^2+\frac{\alpha_1}{4\pi}
M_1^2) + (3 \frac{\alpha_2}{4\pi} + \frac{\alpha_1}{4\pi}) {\mu}^2
- 3 \frac{h_t^2}{(4 \pi)^2}{\mu}^2 + \frac{1}{2} \frac{\alpha_1}{4 \pi}S
\hspace{4.5cm} \;\;
\nonumber\\
&&\frac{dm_2^2}{dt}=(3\frac{\alpha_2}{4\pi}M_2^2+\frac{\alpha_1}{4\pi}
M_1^2) + (3 \frac{\alpha_2}{4\pi} + \frac{\alpha_1}{4\pi}) {\mu}^2
- 3 \frac{h_t^2}{(4 \pi)^2} (m_{Q_L}^2+m_{U_R}^2+m_2^2+A_t^2)
- \frac{1}{2} \frac{\alpha_1}{4\pi} S
\label{dmQdt}
\end{eqnarray}
where we have also included the RGEs of the Higgs doublets $H_1$ and
$H_2$ since they appear in the squark formulae.

Eqs.(\ref{dmQdt}) can be analytically solved and we obtain
(taking into account the
lack of universality in the scalar masses):
\newpage
\begin{eqnarray}
m_{Q_L}^2(t) &=& m_{Q_L}^2(0) + M^2 \big( \frac{1}{3} e(t) +
\frac{8}{3}\frac{\alpha_3(0)}{4\pi}f_3(t)+\frac{\alpha_2(0)}{4\pi}f_2(t)
- \frac{1}{9} \frac{\alpha_1(0)}{4\pi}f_1(t) \big) + \frac{1}{3}AMf(t) -
\nonumber\\
& & \frac{1}{3} k(t) A^2 - \frac{h_t^2(0)}{(4 \pi)^2} \frac{F(t)}{D(t)}
(m_2^2(0)-{\mu}^2(0)+m_{U_R}^2(0)+m_{Q_L}^2(0)) - \frac{1}{6}S(0)f'_1(t)
\nonumber\\
m_{U_R}^2(t) &=& m_{U_R}^2(0)+ M^2 \big( \frac{2}{3} e(t) - H_8(t) \big)
+ \frac{2}{3} A M f(t) - \frac{2}{3} k(t) A^2 -
\nonumber\\
& & 2 \frac{ h_t^2(0)}{(4 \pi)^2} \frac{F(t)}{D(t)} ( m_2^2(0) -
{\mu}^2(0) + m_{U_R}^2(0) +m_{Q_L}^2(0)) + \frac{2}{3} S(0) f'_1(t)
\nonumber\\
m_1^2(t) &=& {\mu}^2(0) l(t) + ( m_1^2(0)-{\mu}^2(0) ) + M^2 g(t)
+ \frac{1}{2} S(0) f'_1(t)
\nonumber\\
m_2^2(t) &=& {\mu}^2(0) l(t) + M^2 e(t) + A M f(t) - k(t) A^2 +
\big( m_2^2(0) - {\mu}^2(0) \big) -
\nonumber\\
& & 3 \frac{h_t^2(0)}{(4\pi)^2} \frac{F(t)}{D(t)} \big( m_2^2(0) -
{\mu}^2(0)+m_{U_R}^2(0)+m_{Q_L}^2(0) \big) - \frac{1}{2} S(0) f'_1(t)
\label{m22t}
\end{eqnarray}
where $f'_1$ was defined in eq.(\ref{ftcb}) and the functions
$l,e,f,k,H_8,f_i,F,D$
are defined in eq.(\ref{dmQdt}) and appendix B of
ref.\cite{ilm}\footnote{They depend not only on $t$ but also on
$\alpha_i(0)$ and $h_t^2(0)$}. For the sake of simplicity we have
taken $M_i(0)=M \;\;i=1,2,3$ in the previous equations but it is
straightforward to extend this study for non--universal gaugino masses.
One has to add to the squarks in eq.(\ref{m22t}) the scalar potential
D--term contribution as in eq.(\ref{mcmm}) and the $(mass)^2$ of the
corresponding fermionic partner.

The analysis of the $U(1)_Y$ D--term contribution to the low--energy
stop and left sbottom masses is more complicated than for the rest of
particles since new parameters appear explicitly in the formulae, in
particular, the top--quark mass $m_t$, the $\mu$ parameter and the
trilinear soft term $A_t(0) \equiv A$. One can eliminate one of them
(e.g. $\mu$, which is the one we know the least) in terms of the others
by imposing appropiate symmetry breaking at the weak scale. We are not
going to review here this well known technology which may be found e.g.
in refs.[7,8] (see \cite{hpn} for reviews). Let us just remark that with
this constraint the vacuum expectation values of the Higgs bosons will
be fixed and therefore the scalar potential D--term contribution will be
as well. The value of the top mass is quite constrained by the LEP and
CDF data so that $m_t$ is not a source of big uncertainty. Now, the
appearence of the $U(1)_Y$ D--term in the above formulae is important
not only because contributes to the physical masses but also because
modifies the electroweak symmetry breaking conditions. Notice that $tan
\beta$ which appears in chargino and neutralino masses and in the ratio
of U-- and D--quark masses depends on $m_{1,2}$. In order to see the
qualitative importance of this contribution to the third generation
squark masses, let us choose\footnote{This is the simplest case in order
to compare the third generation
masses with those of the first and second generation.
Notice that the physical mass contours in the m, M plane will also be
conics for fixed values of $h_t(0)$ (see eq.(\ref{m22t})).} $A=0$ and
for the sake of simplicity we are going to further assume that $B=A-m$.
In fact, the latter assumption has in general a small influence on the
squark spectra. In fig.3 we have plotted as an example different lightest
stop physical mass contours in the m, M plane for $m_t \simeq 145$ GeV
after diagonalizing the mass matrix. The diagonalization
is due to the usual
$t_L-t_R$ mixing terms. They also originate a shift in the fig.3 with
respect to the other generations. We
compare the non--universal case $c^2=-15$ (taking e.g. the simple case
$m_{Q_L}(0)=m_{U_R}(0)=m_{H_{1,2}}(0)=m$) with the universal one $c^2=0$.
The same as the other generations, it is obvious from the figure that the
value of $c^2$ is important to determine the low--energy masses. In fact,
its influence is even larger than for the two first generations (notice
that the physical mass contours behave as hyperbolas for $c^2$ around
--15 instead of --30 as in fig.2), due to the extra contributions
proportional to $m^2$ which appear in eq.(\ref{m22t}) with respect to
eq.(\ref{mcmm}). For $c^2=0$, the region of small M cannot be reached
because the symmetry breaking conditions have no solution. The analysis
for the left sbottom can be carried out in the same way. However, the
influence of $c^2$ is less important than for the lightest stop since
the hypercharge $Y_{b_L}= \frac{1}{6}$ is small and there is no mixing.

Finally, let us show an explicit example where soft masses are not
universal, contrary to the usual assumptions in the MSSM. In string
theory there is no reason to expect soft scalar masses degeneracy to
hold  generically\footnote{However, it is worthy of remark that gaugino
masses are degenerated at $M_{GUT}$ due to the universal contribution of
$S$ to the gauge kinetic function up to small one--loop effects [3-5].}.
In fact, in orbifold constructions \cite{dhvw} the masses depend on the
modular weights of the particles and therefore they show a lack of
universality [3-5]. In particular, assuming vanishing cosmological
constant, the following result was obtained \cite{bim}:
\begin{eqnarray}
m_{\phi}^2 = m^2 \; ( 1 + n_{\phi} {cos}^2 \theta )
\label{mmnc}
\end{eqnarray}
where the modular weights of the matter fields $n_{\phi}$ are normally
negative integers. For example, in the case of $Z_N$ orbifolds the
possible modular weights of matter fields are -1,-2,-3,-4,-5. Fields
belonging to the untwisted sector have $n_{\phi}=-1$. Fields in twisted
sectors of the orbifold but without oscillators have usually modular
weight -2 and those with oscillators have $n_{\phi} \leq -3$ (we
direct the reader to ref.\cite{il1} for an explanation of these points).
The angle parameter $\theta$ says where the source of SUSY--breaking
resides, either predominantly in the dilaton $S$ sector ($sin \theta =1$
limit) or in the modulus $T$($sin\theta =0$ limit). In particular:
$ tan \theta = \frac{\langle F^S \rangle}{\langle F^T \rangle} $, where
$F^S$ is the dilaton auxiliary field (the VEV of $S$ gives the inverse
square of the gauge coupling constant) and $F^T$ is the modulus auxiliary
field (the VEV of $T$ parametrize the size and shape of the compactified
space). Again we direct the reader to ref.\cite{bim} for going into
details.

Notice that only the $sin\theta=1$ limit is universal\footnote{This
limit was obtained in ref.\cite{kl}. However, it is worth noticing
here that in gaugino condensation, the only mechanism so far analysed
capable of generating a hierarchical SUSY breakdown, such limit is not
fulfilled.}. As $sin \theta$ decreases, the modular weight dependence
increases and the resulting soft terms are not universal \cite{bim}. In
particular, fields with higher (negative) modular weight have smaller
soft masses than those with smaller weight. For the masses of scalar
particles with modular weights -1,-2 and -3 eq.(\ref{mmnc}) gives
respectively
\begin{eqnarray}
m_{-1}^2 = m^2 \; sin^2 \theta \; ; \;\;
m_{-2}^2 = m^2 \; (1 - 2 cos^2 \theta) \; ; \;\;
m_{-3}^2 = m^2 \; (1 - 3 cos^2\theta)
\label{m1m2m3}
\end{eqnarray}
This was in fact the case of ref.\cite{ilr} where the following modular
weights for the MSSM fields were obtained (assuming flavour
independence) by imposing the appropiate joining of coupling constants
at $M_{GUT}$:
\begin{eqnarray}
n_{Q_L}=n_{D_R}= -1 \; ; \;\; n_{U_R}=-2 \; ; \;\;
n_{L_L}=n_{E_R}= -3 \; ; \;\; n_{H_1}+n_{H_2}=-4,-5
\label{nqlndr}
\end{eqnarray}
This case shows that, even for large $sin\theta$, the deviation from
universality can be important. Let us consider for example $sin \theta
= 0.82$. Then, from eqs.(\ref{m1m2m3}) and (\ref{nqlndr}) we obtain
\begin{eqnarray}
m_{Q_L,D_R}=0.82 \; m \; ; \;\;
m_{U_R}=0.59 \; m \; ; \;\;
m_{L_L,E_R}=0.13 \; m
\label{m082}
\end{eqnarray}
Thus, the non--universality of soft masses is large and the influence
of the $U(1)_Y$ D--term in the spectrum is crucial. Notice that,
taking for the sake of definiteness $n_{H_2}=-1,n_{H_1}=-3,\;c^2=2.62$
whereas e.g. $c^2_{U_R}=0.34$ and $c^2_{L_L,E_R}=0.02$. This produces a large
effective modification of $U_R,L_L,E_R$ masses at $M_{GUT}$
(see the first term of eq.(\ref{mtcc})).

\vspace{2cm}
\noindent
{\Large\bf{Acknowledgements}}

We thank M. Quir\'os and I. Antoniadis for interesting discussions. We
also thank J.A. Casas and L.E. Ib\'{a}\~nez for useful comments. The work
of A.LL. was supported by the Ministerio de Educaci\'on y Ciencia
through a FPI grant.

\newpage
\noindent
{\Large\bf{Figure captions}}

\vspace{2cm}

FIGURE 1: Allowed values of the soft parameters m, M for different right
selectron ($E_R$) masses. The solid (dotted) lines correspond to the
limit $tan \beta = 1$ ($tan \beta = \infty$). $c^2$ is the parameter
defined in eq.(\ref{cdcd}) and represents the lack of universality of
the soft scalar masses.

FIGURE 2: The same as fig.1 but for the right U--squark ($U_R$).

FIGURE 3: Allowed values of the soft parameters m, M for different
lightest stop masses.

\end{document}